\documentstyle[preprint,aps,prb]{revtex}

\begin{document}

\title{A Variational Sum-Rule Approach to
Collective Excitations of\\ a Trapped Bose-Einstein Condensate}

\author{
Takashi {\sc Kimura}, Hiroki {\sc Saito} and Masahito {\sc Ueda}
}
\address{
Department of Physical Electronics, 
Hiroshima University, Higashi-Hiroshima 739-8527, Japan\\
and Core Research for Evolutional Science and Technology (CREST), 
JST, Japan
}
\date{\today}
\maketitle
\begin{abstract}
It is found that combining 
an excitation-energy sum rule with Fetter's trial 
wave function gives almost exact low-lying collective-mode  
frequencies of a trapped Bose-Einstein condensate at zero temperature.
\end{abstract}
\medskip

%\kword
%{
%Bose-Einstein Condensation, 
%collective excitation, Bogoliubov approximation, 
%sum-rule approach, Fetter's variational wave function 
%}

%\sloppy

%\footnote{
%E-mail : kimura@qnt26.qp.hiroshima-u.ac.jp
%}

%\pacs{PACS numbers: 03.75.Fi, 05.30.jp, 67.40.Db, 67.55.Jd}

Realizations of Bose-Einstein condensation (BEC) in trapped atomic 
gases\cite{Anderson,Bradley,Davis,Bradley2} have enabled us to 
study many-body physics of weakly interacting bosons with 
unprecedented precision.
Experiments on elementary excitations of BEC\cite{Jin,Mewes,Jin2,Kurn} 
present a new challenge because inhomogeneity and
rotational symmetries of the system must be explicitly 
taken into account. 
A number of theoretical studies have been reported in 
literature.\cite{Smerzi,Castin,Feder,Kagan,Edwards,Esry,Stringari,Perez,Hutchinson}
The results of numerical analyses\cite{Edwards} based on the 
Bogoliubov approximation\cite{Bogoliubov,Fetter2}
have so far found the best agreement with 
those of experiments.\cite{Jin}
Stringari\cite{Stringari} obtained exact analytic
expressions of collective-mode frequencies 
in the Thomas-Fermi (TF) limit.\cite{Goldman,Huse,Huse2,Gross,Gross2}
P\'{e}rez-Garc\'{\i}a et al.\cite{Perez} 
assumed a Gaussian trial wave function
which yields correct collective-mode 
frequencies at the weak and strong coupling limits but
 overestimates them at an intermediate regime.
In this Letter, it is found that combining
an excitation-energy sum rule\cite{Wagner} 
with Fetter's trial wave function\cite{Fetter} 
gives the almost exact  frequencies of a trapped Bose system 
at zero temperature.
Our method will be shown to yield
collective-mode frequencies in excellent agreement with experimental 
results as well as numerical ones based on the Bogoliubov approximation.

Consider a Bose system described by the Hamiltonian,
$H=T+U+H_{\rm int}$, where  
$T=-(\hbar^2/2M)\sum_i \nabla_i^2$,
$U=(M/2)\sum_i(\omega_x^2 x_i^2+\omega_y^2 y_i^2+\omega_z^2 z_i^2)$ 
and $H_{\rm int}=(2\pi\hbar^2a/M)\sum_{i\neq j}\delta({\bf r}_i-{\bf r}_j)$
describe the kinetic energy, the confining potential energy and 
the inter-particle interaction energy, respectively. 
Here $M$ and $a$ denote the atomic mass 
and the $s$-wave scattering length, respectively, and 
$i$ denotes the particle index. 
Let $\{|n\rangle\}$ be a complete set of exact eigenstates of $H$ 
with eigenvalues $\{E_n\}$, where $n$ represents a complete set of quantum
numbers that uniquely specify the many-body state of our system.
For brevity of notation, we assume $n$ to be nonnegative integers such 
that $E_0\le E_1\le E_2\le \cdot \cdot \cdot$. 
We are interested in low-lying excitations  of  many-body states 
which are excited by a general excitation operator $F$. 
Let $|0\rangle$ be the ground state and 
$|1\rangle$ be the lowest excited state with the excitation energy
given by $\hbar\omega_{10}=E_1-E_0$.
Because $\omega_{10}$ is  not greater than 
$\hbar\omega_{n0}\equiv E_n-E_0$ for $n\ge 2$, 
we obtain the following inequality
\begin{eqnarray}
\omega_{10}^2\leq\omega_{10}^2
\frac{|\langle 1|F|0\rangle|^2+\sum_{n\neq 1}
|\langle n|F|0\rangle|^2\left(\frac{\omega_{n0}}{\omega_{10}}\right)^3}
{|\langle 1|F|0\rangle|^2+\sum_{n\neq 1}|\langle n|F|0\rangle|^2
\frac{\omega_{n0}}{\omega_{10}}},\label{sumrule}
\label{ineq1}
\end{eqnarray}
which shows that an upper bound $\hbar\omega^{\rm upper}$ of the lowest
excitation energy is given by 
\begin{eqnarray} 
\hbar\omega^{\rm upper}=\sqrt{m_3/m_1},\label{eq2}
\end{eqnarray}   
where $m_p\equiv\sum_n |\langle 0|F|n\rangle |^2(\hbar\omega_{n0})^p$ 
is the $p$-th moment of the excitation energy.\cite{Stringari,Wagner}
Similar methods which evaluate the excitation energy have
been used extensively 
in the field of nuclear physics.\cite{Bohigas}
The advantage of this formula is 
that $m_1$ and $m_3$ can be expressed 
as expectation values of commutators 
between $F$ and $H$ with respect to the ground state $|0\rangle$ as 
$m_1=\frac{1}{2}\langle0|\big[F^\dagger,[H,F]\big]|0\rangle$ 
and 
$m_3=\frac{1}{2}\langle0|[[F^\dagger,H],\big[H,[H,F]\big]]|0\rangle$,  
and we can therefore find $\hbar\omega^{\rm upper}$
without the necessity of finding excited states.\cite{Stringari} 
To obtain $\omega^{\rm upper}$ which is very close to $\omega_{10}$,
we need to find a correct excitation operator $F$ and 
an accurate ground-state wave function. 

We first consider the case of an axially symmetric trap
($\omega_x=\omega_y\equiv\omega_\perp$), which is relevant 
to recent experiments.\cite{Jin,Mewes,Jin2,Kurn} 
Fetter proposed a trial wave function for the condensate with repulsive 
 interaction as\cite{Fetter}
\begin{eqnarray}
\Psi({\bf r})=c_0 \Big(1-\frac{r_\perp^2}{d^2_\perp R^2_\perp}
                    -\frac{z^2}{d^2_z R^2_z}\Big)^{(1+\eta)/2},\label{wave}
\end{eqnarray}
if $\frac{r_\perp^2}{d^2_\perp R^2_\perp}+\frac{z^2}{d^2_z R^2_z}\le 1$ 
and zero otherwise, where $r_\perp\equiv\sqrt{x^2+y^2}$, 
$c_0$ is the normalization constant, 
$d_j\equiv\sqrt{\hbar/M\omega_j} (j=\perp,z)$ is the oscillator 
length, and $R_j$ and $\eta$ are the variational parameters which 
are determined so as to minimize the total energy 
$E_0=\langle T\rangle+\langle U\rangle +\langle H_{\rm int}\rangle$.  
Fetter showed that the trial wave function (\ref{wave}) 
smoothly interpolates between the noninteracting limit 
($\Psi({\bf r})\propto{\rm exp}(-r_\perp^2/2d_\perp^2-z_\perp^2/2d_z^2)$
for $\eta\sim\sqrt{R_\perp},\sqrt{R_z}\rightarrow\infty$) 
and the strongly interacting limit ($\Psi({\bf r})\sim c_0 
(1-r_\perp^2/d_\perp^2R_\perp^2-z^2/d_z^2R_z^2)^{1/2}$
for $\eta\rightarrow 0$). 
However, how well it describes an intermediate regime had not been  
investigated until now. 
We therefore examine the accuracy of Fetter's variational wave function 
for various atomic numbers $N_0$. 
Table I compares the expectation values of the total, kinetic, potential, 
and interaction energies obtained by Fetter's 
wave function with those obtained numerically 
according to the method reported in ref. 28.
The agreement is excellent and we thus conclude that eq. (3) serves 
our purpose very well. 

Experimentally, low-lying collective modes are excited by superimposing 
small ac currents with appropriate phase relationship through magnetic coils.
This leads to small modulations of the frequencies of the confining potential. 
We consider $|m|=2$ and $m=0$ modes 
which are relevant to recent experiments.\cite{Jin,Mewes,Jin2,Kurn} 
The collective mode with magnetic quantum number 
$|m|=2$ describes the situation 
in which the condensate expands in one direction and simultaneously 
contracts in the other, thus maintaining its volume. 
This mode can be excited by modulating two radial trap frequencies  
out of phase but by the same amount $\delta\omega \ll \omega_\perp$.
The resulting perturbation which defines the excitation operator is given 
by $F_{|m|=2}=M/2\sum_i\big\{
[(\omega_\perp+\delta\omega)^2-\omega_\perp^2]\ x_i^2
+\ [(\omega_\perp-\delta\omega)^2-\omega_\perp^2]\ y_i^2\big\}
\approx M\omega_\perp\delta\omega\sum_i(x_i^2-y_i^2)$. We  note that 
$F$ is proportional to $\sum_ir_i^2(Y_{2,2}+Y_{2,-2})$, 
where $Y_{lm}$ is the spherical harmonic function.\cite{Stringari}
We can therefore take the excitation operator as 
$F_{|m|=2}=\sum_i(x_i^2-y_i^2)$, 
where the numerical factors are dropped because they are 
canceled in forming the ratio (2).
The commutator $\big[H,F_{|m|=2}\big]$ is calculated as 
\begin{eqnarray}
\big[H,F_{|m|=2}\big]&=&\frac{1}{2M}\sum_i[p_{xi}^2+p_{yi}^2, 
x_i^2-y_i^2]\nonumber\\
&=&-\frac{2i\hbar}{M}\sum_i(x_i p_{xi}-y_i p_{yi}).\label{ap1}
\end{eqnarray}
We obtain the first moment $m_1$ as 
\begin{eqnarray} 
m_1&=&\frac{1}{2}\Big\langle \Big[F_{|m|=2}^\dagger,
\big[H,F_{|m|=2}\big]\Big]\Big\rangle \nonumber\\
&=&\frac{2\hbar^2}{M}\Big\langle\sum_i(x_i^2+y_i^2)\Big\rangle
=\frac{8\hbar^2}{M^2\omega_\perp^2}\langle U_\perp \rangle,
\label{ap2}
\end{eqnarray}
where $U_\perp$ is a radial ($x$ or $y$) component 
of the trap potential energy
and $\langle \cdot\cdot\cdot\rangle$ denotes the expectation value 
over the trial wave function (\ref{wave}).  
We calculate the third moment $m_3$ as
\begin{eqnarray} 
m_3&=&\frac{1}{2}\Big\langle\big[\big[F_{|m|=2}^\dagger,H\big],
\Big[H,\big[H,F_{|m|=2}\big]\Big]\big]\Big\rangle\nonumber\\
&=&\frac{16\hbar^4}{M^2}(\langle T_\perp\rangle +\langle U_\perp \rangle),
\end{eqnarray}
where $T_\perp$ is a radial ($x$ or $y$) component 
of the kinetic energy.
Because the upper-bound frequency does not directly 
depend on the interaction energy $\langle H_{\rm int}\rangle$,  
we obtain the upper-bound frequency as
\begin{eqnarray} 
\omega^{\rm upper}(m=2)=\omega_\perp 
\sqrt{2(1+\langle T_\perp\rangle/\langle U_\perp\rangle)}.\label{eq4}
\end{eqnarray} 
In the absence of inter-particle interaction 
we find that $\langle T_\perp\rangle=\langle U_\perp\rangle$, 
so that   
$\omega^{{\rm upper}}(m=2)=2\omega_\perp$, while in the  TF limit 
we have $\langle T_\perp\rangle=0$, so that eq. (\ref{eq4})
reduces to the exact result 
$\omega^{{\rm upper}}(m=2)=\sqrt{2}\omega_\perp$.\cite{Stringari}

The $m=0$  mode describes an excitation in which the 
condensate alternately expands and contracts in the radial direction. 
This part of the excitation is described by  
$F_{m=0}=M/2\sum_i\big\{
[(\omega_\perp+\delta\omega)^2-\omega_\perp^2]r_{\perp i}^2\big\}
\approx M\omega_\perp\delta\omega\sum_i r_{\perp i}^2$.\cite{Stringari}
Because of the repulsive interaction,
however, the condensate should also undergo oscillations in 
the axial direction which must be out of phase 
with the radial motion.\cite{Jin}
Hence we consider the excitation operator of the form 
$F_{m=0}=M\omega_\perp\delta\omega\sum_i(r_{\perp i}^2-\alpha z_i^2),$
where $\alpha$ is another variational parameter. 
We note that $F$ is a linear 
combination of two modes $\sum_ir_i^2$ ($n=1,l=m=0$) and 
$\sum_ir_i^2Y_{2,0}(\theta_i,\phi_i)$ ($n=0,l=2,m=0$). 
The excitation operator for the $m=0$ mode can therefore be taken as 
$F_{m=0}=\sum_i(x_i^2+y_i^2-\alpha z_i^2)$ for simplicity
as in the $|m|=2$ case.
The first moment $m_1$ is calculated as,
\begin{eqnarray}
m_1=\frac{4\hbar^2}{M^2}\Big(2\frac{\langle U_\perp \rangle}{\omega_\perp^2}
+\alpha^2\frac{\langle U_z \rangle}{\omega_z^2}\Big),
\end{eqnarray}
where $U_z$
is the axial component of the trap potential energy.
The third moment $m_3$ is calculated as
\begin{eqnarray}
m_3=\frac{8\hbar^4}{M^2}\langle[2(T_\perp+2U_\perp)+\alpha^2(T_z+U_z)
+(1-\frac{\alpha}{2})^2E_{\rm int}]\rangle\label{ap4}
\end{eqnarray}
where $T_z$
is the axial component of the kinetic energy.
We thus obtain the upper-bound frequency as
\begin{eqnarray}
\omega^{\rm upper}(m=0,\alpha)=
\Big[2\frac{2(\langle T_\perp\rangle+
\langle U_\perp\rangle)+\alpha^2(\langle T_z\rangle+\langle U_z\rangle)
+(1-\alpha/2)^2\langle H_{\rm int}\rangle}
{2\langle U_\perp\rangle/\omega_\perp^2+\alpha^2\langle U_z
\rangle/\omega_z^2}\Big]^{\frac{1}{2}}.
\end{eqnarray}
By minimizing $\omega^{\rm upper}(m=0,\alpha)$ 
with respect to $\alpha$, 
we find $\omega^{{\rm upper}}(m=0)=2\omega_\perp$ 
in the noninteracting limit and 
$\omega^{{\rm upper}}(m=0)=\omega_\perp
(2+\frac{3}{2}\lambda^2-\frac{1}{2}\sqrt{9\lambda^4-
16\lambda^2+16})^{\frac{1}{2}}$
($\lambda\equiv\omega_z/\omega_\perp$) in the strong interacting limit. 
This latter result is identical 
to that is obtained in ref.$\! 15$
by another method.
The agreement shows that the excitation operator $F_{m=0}$
that we chose is indeed correct. 
We also note that if 
$\omega^{{\rm upper}}$ is maximized with respect to $\alpha$, 
we obtain $\omega^{{\rm upper}}(m=0)=\omega_\perp 
(2+\frac{3}{2}\lambda^2+\frac{1}{2}\sqrt{9\lambda^4-
16\lambda^2+16})^{\frac{1}{2}}$, 
which also coincides with the result reported in ref. 15
with higher frequency. Because states excited by 
$F_{m=0}$ are restricted to 
states which are constructed by 
linear combinations of 
the $n=1,l=m=0$ mode and the $n=0,l=2,m=0$ mode, there 
should be two values of $\alpha$ that make $\omega^{{\rm upper}}$ extremal 
and the corresponding states should describe the two lowest-energy 
excited states as obtained above.

Our method can also be applied to the dipole ($l=1$) 
modes \cite{Stringari}, 
which correspond to the center-of-mass motion of the condensate,  
and should therefore not be affected by the inter-particle interaction. 
This is known as the generalized Kohn's theorem.
We consider the excitation operator 
$F=M/2\sum_i\omega_\perp^2[(x_i+\delta)^2-x_i^2]
\propto\sum_ix_i \propto \sum_ir_i(Y_{1,1}({\bf r}_i)-Y_{1,-1}({\bf r}_i))$ 
or $F=M/2\sum_i\omega_\perp^2[(y_i+\delta)^2-y_i^2]
\propto\sum_i r_i(Y_{1,1}({\bf r}_i)+Y_{1,1}({\bf r}_i))$ for $|m|=1$, 
and $F=M/2\sum_i\omega_z^2[(z_i+\delta)^2-z_i^2]
\propto\sum_iz_i\propto\sum_ir_iY_{10}({\bf r}_i)$ for $m=0$.  
Substituting $F$ into Eq. (\ref{eq2}), we easily obtain 
$\omega^{{\rm upper}}(l=1,|m|=1)=\omega_\perp$
and $\omega^{{\rm upper}}(l=1,m=0)=\omega_z$.
We thus find that the dipole-mode frequencies
obtained by our method coincide with trap frequencies,
being independent of the strength of interaction.

Figure 1 compares our analytical results (solid curves) 
with the experimental data (dots)
taken from ref. 5 for ${}^{87}$Rb atoms, where  
we use the same parameters as in the experiment; 
$a=109a_0$ ($a_0$ is the Bohr radius) and  
$\omega_z/\sqrt{8}=\omega_\perp=2\pi\times132$Hz. 
Our results for both $m=0$ and $|m|=2$ modes 
are in excellent agreement with those of the experiment and 
with those obtained with the Bogoliubov approximation\cite{Edwards}. 
We have also calculated the upper-bound frequencies 
using numerically calculated values of $\langle T\rangle$ and 
$\langle U\rangle$ for
4500 atoms and obtained 
$\omega^{\rm upper}(|m|=2)=1.454\omega_\perp$ and
$\omega^{\rm upper}(m=0)=1.871\omega_\perp$. 
These results are in excellent agreement with those obtained 
using Fetter's variational wave function, that is, 
$\omega^{\rm upper}(|m|=2)=1.450\omega_\perp$ and 
$\omega^{\rm upper}(m=0)=1.870\omega_\perp$.  

We briefly describe the results of our study for 
the case of a spherically symmetric 
trap ($\omega_\perp=\omega_z\equiv\omega_0$),
where we can compare our results with those 
reported in ref. 15 and with 
numerical ones based on the Bogoliubov approximation\cite{Hutchinson}.
For the quadrupole mode ($l=2$), 
we consider the case of  $m=2$ without loss of generality, where 
$F=M\delta\omega\sum_i(x_i^2-y_i^2)$, and find that 
$\omega^{{\rm upper}}(l=0)=\omega_0\sqrt{2+\langle T\rangle/\langle U\rangle}$.
For the monopole mode ($n=1,l=0$), where  
$F=M\delta\omega\sum_ir_i^2$, we find that 
$\omega^{{\rm upper}}(l=0)=\omega_0
\sqrt{5-\langle T\rangle/\langle U\rangle}$, 
where we have used the virial identity
$2\langle T\rangle-2\langle U\rangle+3\langle H_{\rm int}\rangle =0$.
These results are the same as those reported in ref. 15, 
although analytic evaluation of $\langle T\rangle$ and $\langle U\rangle$ 
in an intermediate strength of interaction is not so far available from
the TF limit. 

In Fig. 2, we present our results for both modes 
with the same parameters ($\omega_0=2\pi\times200$Hz, $a=110a_0$)  
as those reported in ref. 17
using the Bogoliubov approximation. 
Our results are almost indistinguishable from those 
obtained numerically using the Bogoliubov approximation.

In conclusion, we have presented an analytical method 
to evaluate almost exactly 
the collective-mode frequencies of a trapped Bose gas. 
The results are in excellent agreement with those of 
the Bogoliubov approximation\cite{Edwards} and those of 
the experiments\cite{Jin} at zero temperature. 
%%%%%%%%%%%%%%%%%%%% References %%%%%%%%%%%%%%%%%%%%%%%%%%%%%%%%%%%%%%%%

%%%%%%%%%%%%%%%%%%%% Figure Captions %%%%%%%%%%%%%%%%%%%%%%%%%%%%%%%%%%%
\begin{table}
\caption{Comparison of total energy $\langle E_{\rm tot}\rangle$, 
kinetic energy $\langle T\rangle$, confining-potential energy 
$\langle U\rangle$ and 
interaction energy $\langle H_{\rm int}\rangle$ obtained 
using Fetter's variational wave function with those obtained 
numerically according to ref. 28 (in parentheses) for an axially-symmetric trap. 
We take $\sqrt{8}\omega_\perp=\omega_z=2\pi\times220$ Hz and $a=100a_0$.  
Energies are shown in units of $\hbar\omega_\perp$ and 
$N_0$ is the atom number. }
\label{table:1}
\begin{tabular}{@{\hspace{\tabcolsep}\extracolsep{\fill}}cccccc} \hline
$N_0$ & $\langle E_{\rm tot}\rangle$  & $\langle T\rangle$
& $\langle U\rangle$ & $\langle H_{\rm int}\rangle$ \\ \hline  
%\begin{tabular}{|c|c|c|c|c|}\hline
%\makebox[10mm]{$N_0$} &
%\makebox[14mm]{$\langle E_{\rm tot}\rangle$} &
%\makebox[14mm]{$\langle T\rangle$} & 
%\makebox[14mm]{$\langle U\rangle$} &
%\makebox[14mm]{$\langle H_{\rm int}\rangle$}
1000 & 3.86 (3.84) & 0.75 (0.76) & 2.17 (2.15) & 0.95 (0.93) \\ 
10000 & 7.83 (7.76) & 0.46 (0.45) & 4.61 (4.57) & 2.76 (2.74)\\ 
20000 & 10.06 (9.98) & 0.40 (0.38) & 5.95 (5.91) & 3.70 (3.68) \\ \hline
\end{tabular}
\end{table}

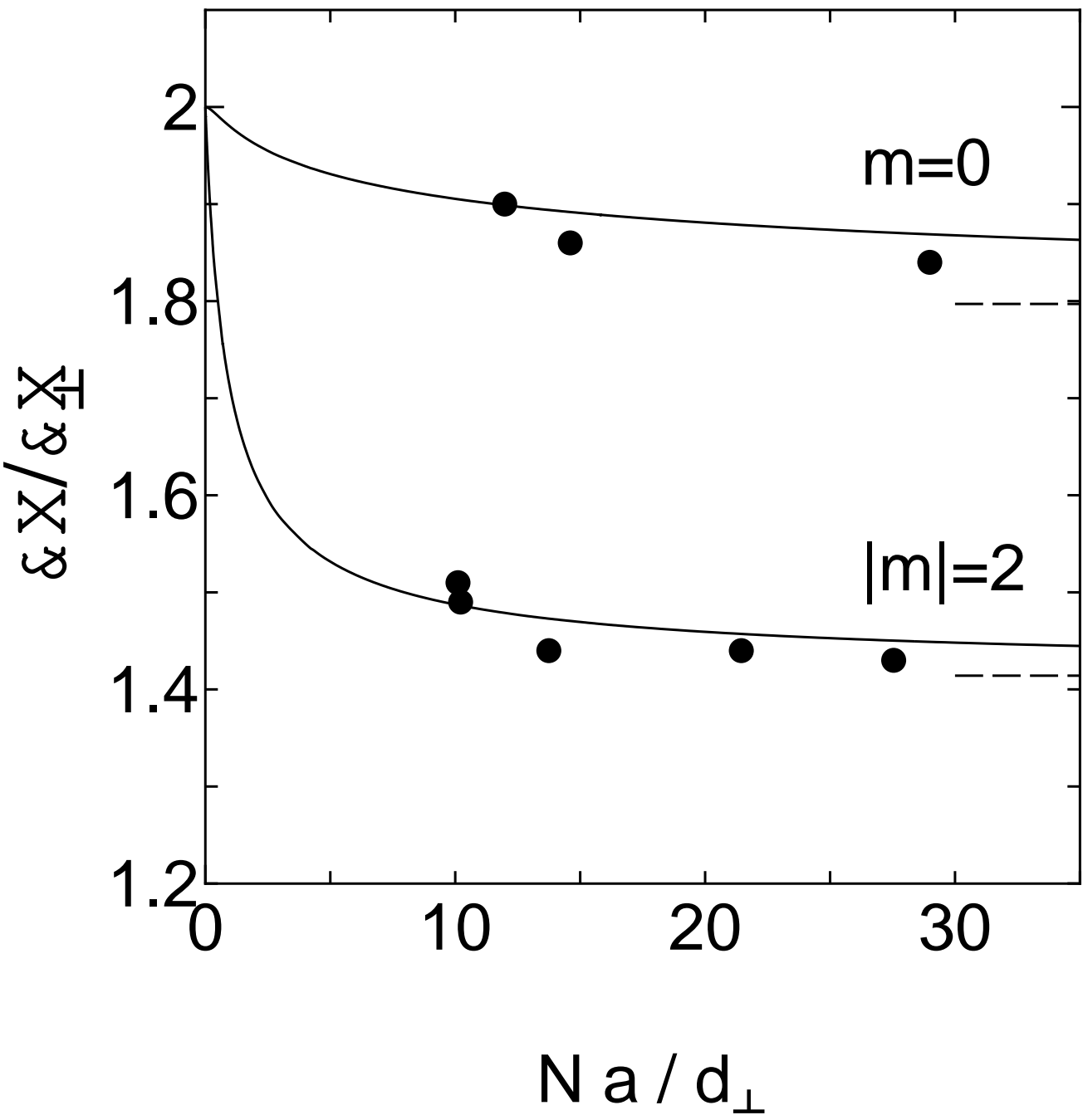
\begin{figure}
\begin{center}
%\epsfigure{file=figure1.eps,height=7cm}
\caption{
Collective-mode frequencies 
of the $|m|=2$ and the $m=0$ modes 
in an axially symmetric trap, where we assume
$\omega_z/\protect\sqrt{8}=\omega_\perp=2\pi\times132$Hz and $a=109a_0$. 
The solid curves show our results, 
the dots show the experimental data reported in ref. 5, and the
dashed lines show the results in ref. 15.
}
\end{center}
\label{axi}
\end{figure}

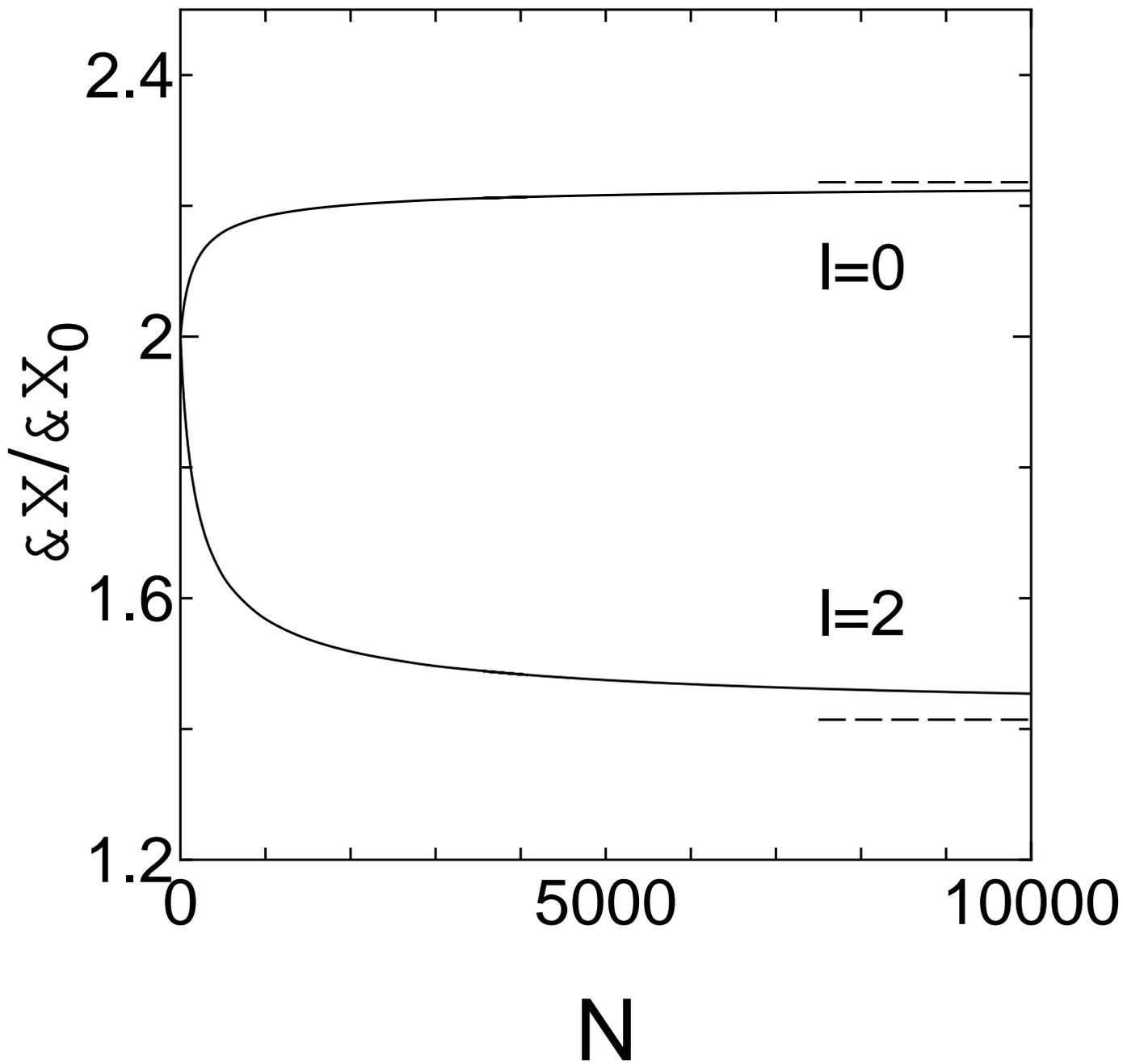
\begin{figure}
\begin{center}
%\epsfigure{file=figure2.eps,scale=0.3}
\caption{
Collective-mode frequencies 
for the quadrupole and the monopole modes 
in a spherically symmetric trap. The dashed lines show the 
results reported in ref. 17. 
}
\end{center}
\label{sph}
\end{figure}\end{document}